\documentclass[aps,prc,twocolumn,showpacs,superscriptaddress,amsmath,amssymb]{revtex4}

\usepackage{graphicx, latexsym, amssymb, amsmath, color, multirow, mathrsfs, CJK, ifpdf}
\usepackage{amsmath}
\usepackage{amsfonts}
\usepackage{amssymb}
\usepackage{bm}
\usepackage{graphicx}
\usepackage{color} 
\usepackage{txfonts}
\usepackage[scaled=1.0]{helvet}
\usepackage[T1]{fontenc}
\usepackage{textcomp}

\makeatletter

\newcommand{\Rmnum}[1]{\expandafter\@slowromancap\romannumeral#1@}
\makeatother

\begin{document}


\title{Toward extremes of angular momentum: \\
Application of the Pfaffian algorithm in realistic calculations}

\author{Long-Jun Wang}
\affiliation{Department of Physics and Astronomy, Shanghai Jiao Tong
University, Shanghai 200240, People's Republic of China}
\author{Fang-Qi Chen}
\affiliation{Department of Physics and Astronomy, Shanghai Jiao Tong
University, Shanghai 200240, People's Republic of China}
\affiliation{China Institute of Atomic Energy, Beijing 102413,
People's Republic of China}
\author{Takahiro Mizusaki}
\affiliation{Institute of Natural Sciences, Senshu University, 3-8-1
Kanda-Jinbocho, Chiyoda-ku, Tokyo 101-8425, Japan}
\author{Makito Oi}
\affiliation{Institute of Natural Sciences, Senshu University, 3-8-1
Kanda-Jinbocho, Chiyoda-ku, Tokyo 101-8425, Japan}
\author{Yang Sun}
\email{sunyang@sjtu.edu.cn} \affiliation{Department of Physics and
Astronomy, Shanghai Jiao Tong University, Shanghai 200240, People's
Republic of China} \affiliation{Institute of Modern Physics, Chinese
Academy of Sciences, Lanzhou 730000, People's Republic of China}

\date{\today}
\begin{abstract}
In a calculation of rotated matrix elements with angular momentum
projection, the generalized Wick's theorem may encounter a practical
problem of combinatorial complexity when the configurations have
more than four quasi-particles (qps). The problem can be solved by
employing the Pfaffian algorithm generally applicable to
calculations of matrix elements for Hartree-Fock-Bogoliubov states
with any number of qps. This breakthrough in many-body techniques
enables studies of high-spin states in a shell-model framework. As
the first application of the Pfaffian algorithm, the configuration
space of the Projected Shell Model is expanded to include 6-qp
states for both positive and negative parities. Taking $^{166}$Hf as
an example, we show that 6-qp states become the main configuration
of the yrast band beyond spin $I \approx 34\hbar$, which explains
the observed third back-bending in moment of inertia. Structures of
multi-qp high-$K$ isomers in $^{176}$Hf are analyzed as another
example.
\end{abstract}


\maketitle

All nucleons of even-even nuclei couple pairwise in their ground
state. Nuclear rotation brings an effect into the system that tends
to break the nucleon pairs. This effect due to the rotation of
nuclei was suggested in 1960 \cite{Mottelson-Valatin} in analogy to
the electron pair breaking in superconductivity due to the existence
of external magnetic fields. However, it was soon after realized
that a sharp phase transition with pair collapse does not occur in
nuclear systems due to the fact that nuclei have a finite size
\cite{Ring}. Another reason is that nucleons have an orbital angular
momentum in addition to spin. The nucleons in the vicinity of the
Fermi surfaces belong to subshells with rather different $j$-values,
and therefore, they feel the Coriolis force very differently. Pairs
in those orbitals with the highest angular momentum $j$, as for
instance the neutron $i_{13/2}$ shell in the rare earth region, feel
a strong Coriolis force, and therefore, break first (the
Stephens-Simon effect \cite{Stephens-Simon}). They contribute to
formation of 2-quasiparticle (qp) state as the main configuration of
the yrast state. The Stephens-Simon effect \cite{Stephens-Simon}
successfully explained the experimental observations of backbending
in moment of inertia for rotating nuclei \cite{First-Backbending}.
As a nucleus rotates faster and faster, subsequent pair-breakings
can occur for the pairs from the next highest $j$ orbitals. In the
rare earth region, proton pairs in the $h_{11/2}$ shell are expected
to break next, which was observed experimentally
\cite{Second-Backbending}. Measurements for further pair breakings
at higher angular momenta are difficult. However, there have been
early \cite{Third-Backbending} and recent evidences
\cite{Hf166-EPJA2000, Hf168PRC2009} of breaking of three nucleon
pairs, which form 6-qp states as the main configuration of the yrast
sequence.

Pair-breaking in nuclei can lead to formation of another special
group of excited states: nuclear isomers. In a deformed potential a
$j$-shell splits up into $2j+1$ $K$-states. Two or more states with
high $K$ quantum numbers can couple to form a high-$K$ multi-qp
configuration. The selection rules of electromagnetic decay hinder
transitions from a high-$K$ state to normal (low-$K$) states,
resulting in a long-lived $K$ isomer \cite{Walker1999Nature}.
Understanding nuclear isomeric states is one of the current topics
in nuclear structure and nuclear astrophysics, and also in potential
applications \cite{Walker1999Nature, Sun2005Nat-Phys,
Walker2005PhysTod, Xu2004PRL, Herzberg2006Nature}. The current
experiments are able to find evidences for 6- or 8-qp isomers (see,
for example, Refs. \cite{Purry95,Dra05}), and discoveries of more
high-$K$ isomers are expected from modern facilities such as the
storage ring \cite{Reed10}. These data provide us with valuable
information on the single-particle structure in deformed potentials.

On the theoretical side, it is a challenge to describe these
phenomena in a shell-model framework if high-order qp states are
involved. A theoretical problem lies in the procedure of computing
the overlap matrix elements \cite{Ring2004many-body} between
arbitrary multi-qp Hartree-Fock-Bogoliubov (HFB) states. The problem
is known as follows. To study heavy, deformed nuclei, a practical
way to build the model space is to start from a deformed
single-particle basis (e.g. the solutions from a HFB or simply from
the Nilsson+BCS method), and perform angular-momentum projection for
the deformed qp states. Shell model diagonalization with a two-body
Hamiltonian is then carried out in the projected multi-qp space.
This is the basic concept of the Projected Shell Model (PSM)
\cite{PSM-review}, which has been extensively applied to the
structure study. However, since the involved overlap matrix elements
of multi-qp states are usually calculated with the generalized
Wick's theorem \cite{PSM-review}, one may encounter a practical
problem of combinatorial complexity when more than 4-qp states are
included in the basis configurations. For example, as many as
hundreds (thousands) terms are to be considered to express each
matrix element with 4-qp (6-qp) state. Therefore, up to recently,
3-qp (4-qp) states are selected among multi-qp configurations that
can practically be treated in the PSM, for odd-mass (even-even)
systems \cite{Chen2012PRC,Liu2011NPA}.

To push the calculation further toward extremes of angular momentum
involving higher order of qp states, a breakthrough in computational
many-body techniques is needed. In nuclear structure physics, the
Pfaffian concept has been introduced \cite{Robledo2009PRC} as a key
mathematical tool for solving the long-standing problem in the phase
determination of the Onishi formula \cite{Onishi1996NP}. Moreover,
it has been shown that the Pfaffian algorithm is very efficient also
for calculating overlap matrix elements \cite{Robledo2011PRC,
Robledo2012PRL, Bender2012PRC, Oi-Misuzaki2012PLB,
Misuzaki-Oi2012PLB, Hu2013}. In particular, by means of Fermion
coherent states and Grassmann integral, some of the present authors
have derived an alternative approach to calculate the rotated matrix
element for general qp states \cite{Mizusaki2013PLB}, which serves
as a theoretical framework to extend the PSM model space.

Let us explain how the Pfaffian algorithm is applied to the PSM. The
PSM employs the deformed Nilsson model \cite{Nilsson1969} to
generate a single-particle basis. Pairing correlations are
incorporated into the Nilsson states by a BCS calculation. The
Nilsson-BCS calculation defines a deformed qp basis from which the
PSM model space is constructed. Three major harmonic-oscillator
shells are taken in the calculation with $N=4,5,6$ ($N=3,4,5$) for
neutrons (protons). Let us use $a^\dag_\nu, a^\dag_\pi$ ($a_\nu,
a_\pi$) to denote neutron and proton qp creation (annihilation)
operators associated with the qp vacuum $|\Phi\rangle$. The multi-qp
configurations up to 6-qp states for even-even nuclei are given as
follows:
\begin{align}\label{Eq.config}
  \Big\{\ & |\Phi\rangle, \ a^\dag_{\nu_i}a^\dag_{\nu_j}|\Phi\rangle, \ a^\dag_{\pi_i}a^\dag_{\pi_j}|\Phi\rangle, \ a^\dag_{\nu_i}a^\dag_{\nu_j}a^\dag_{\pi_k}a^\dag_{\pi_l}|\Phi\rangle, \nonumber \\
& \ a^\dag_{\nu_i}a^\dag_{\nu_j}a^\dag_{\nu_k}a^\dag_{\nu_l}|\Phi\rangle, \ a^\dag_{\pi_i}a^\dag_{\pi_j}a^\dag_{\pi_k}a^\dag_{\pi_l}|\Phi\rangle, \nonumber \\
& \ a^\dag_{\nu_i}a^\dag_{\nu_j}a^\dag_{\nu_k}a^\dag_{\nu_l}a^\dag_{\nu_m}a^\dag_{\nu_n}|\Phi\rangle, \ a^\dag_{\pi_i}a^\dag_{\pi_j}a^\dag_{\pi_k}a^\dag_{\pi_l}a^\dag_{\pi_m}a^\dag_{\pi_n}|\Phi\rangle, \nonumber \\
& \ a^\dag_{\pi_i}a^\dag_{\pi_j}a^\dag_{\nu_k}a^\dag_{\nu_l}a^\dag_{\nu_m}a^\dag_{\nu_n}|\Phi\rangle, \ a^\dag_{\nu_i}a^\dag_{\nu_j}a^\dag_{\pi_k}a^\dag_{\pi_l}a^\dag_{\pi_m}a^\dag_{\pi_n}|\Phi\rangle.  \Big\}
\end{align}
where those 4-qp and 6-qp states from the fifth to tenth items in
(\ref{Eq.config}) are the new configurations introduced in the
present work. Note that the PSM works with multiple
harmonic-oscillator shells for both neutrons and protons. There are
no restrictions in taking the indices $\nu$ (for neutrons) and $\pi$
(for protons) in (\ref{Eq.config}); for example, a 2-qp state can be
of positive parity if both quasiparticles $i$ and $j$ are from the
major $N$-shells that differ in $N$ by $\Delta N = 0,2,\ldots,$ or
of negative parity if $i$ and $j$ are from those $N$-shells that
differ by $\Delta N=1,3,\ldots$.

The PSM wave function is a linear combination of projected states
\begin{align}\label{Eq.wave-func}
|\Psi^\sigma_{IM}\rangle = \sum_{K\kappa} f^\sigma_{IK_\kappa} \hat
P^I_{MK} |\Phi_\kappa\rangle ,
\end{align}
where $|\Phi_\kappa\rangle$ are the multi-qp-states in
(\ref{Eq.config}). $\hat P^I_{MK}$ is the angular momentum
projection operator \cite{Ring2004many-body}
\begin{align}\label{Eq.projec-op}
  \hat P^I_{MK} = \frac{2I+1}{8\pi^2} \int d\Omega D^I_{MK}(\Omega)\hat R(\Omega) ,
\end{align}
with $D^I_{MK}$ being the $D$-function \cite{Angular-book}, $\hat R$
the rotation operator, and $\Omega$ the Euler angle. The energies
and wave functions are obtained by solving the eigenvalue equation:
\begin{align}\label{Eq.diag}
  \sum_{K'\kappa'} \Big(H^I_{K\kappa,K'\kappa'} - E^\sigma_I N^I_{K\kappa,K'\kappa'} \Big) f^\sigma_{IK'_{\kappa'}} =0,
\end{align}
where $H^I_{K\kappa,K'\kappa'}$ and $N^I_{K\kappa,K'\kappa'}$ are
respectively the projected matrix elements of the Hamiltonian and
the norm
\begin{align}
  H^I_{K\kappa,K'\kappa'} = \langle \Phi_\kappa| \hat H \hat P^I_{KK'} |\Phi_{\kappa'}\rangle, \ \ \
  N^I_{K\kappa,K'\kappa'} = \langle \Phi_\kappa| \hat P^I_{KK'} |\Phi_{\kappa'}\rangle .
\end{align}
The central task in numerical calculations is to evaluate rotated
matrix elements in the Hamiltonian and the norm
\begin{align}
  \mathcal{H}_{\kappa\kappa'} = \langle \Phi_\kappa| \hat H [\Omega]
  |\Phi_{\kappa'}\rangle, \ \ \
  \mathcal{N}_{\kappa\kappa'} = \langle \Phi_\kappa| [\Omega] |\Phi_{\kappa'}\rangle
  ,
\end{align}
with the operator $[\Omega]$ defined as \cite{PSM-review}
\begin{align}
  [\Omega] = \frac{\hat R(\Omega)}{\langle \Phi| \hat R(\Omega) |\Phi\rangle}.
\end{align}
$\mathcal{H}_{\kappa\kappa'}$ can be decomposed into terms of the
``linked" contraction and $\mathcal{N}_{\kappa\kappa'}$ (see
Appendix A.3 of Ref. \cite{PSM-review} for details). An explicit
example is for the evaluation of one-body operator $\hat O$,
expressed as
\begin{align}\label{Eq.onebody}
  \langle \Phi_\kappa|\hat O & [\Omega] |\Phi_{\kappa'}\rangle = \langle\Phi| \hat O[\Omega] |\Phi\rangle \langle\Phi_\kappa| [\Omega]|\Phi_{\kappa'}\rangle \nonumber \\
 & + \sum_{ij}(\pm) \Big(\hat O[\Omega]a^\dag_i a^\dag_j \Big) \langle\Phi_{\kappa}| [\Omega]|\Phi_{\kappa'(ij)}\rangle \nonumber \\
 & + \sum_{ij}(\pm) \Big(a_i\hat O[\Omega]a^\dag_j \Big) \langle\Phi_{\kappa(i)}| [\Omega]|\Phi_{\kappa'(j)}\rangle \nonumber \\
 & + \sum_{ij}(\pm) \Big(a_i a_j\hat O[\Omega] \Big) \langle\Phi_{\kappa(ij)}| [\Omega]|\Phi_{\kappa'}\rangle
 ,
\end{align}
where $(\pm)$ is the parity of permutation. In Eq.
(\ref{Eq.onebody}), a state such as $|\Phi_{\kappa'(ij)}\rangle$
means a one obtained by removing qps $a^\dag_{i}$ and $a^\dag_{j}$
from the state $|\Phi_{\kappa'}\rangle$ to 
the ``linked'' contraction (expressed in round parenthese $\Big(\hat
O[\Omega]a^\dag_i a^\dag_j\Big)$, which can be easily evaluated
\cite{PSM-review}). Thus, the main task concentrates on treating
$\mathcal N_{\kappa\kappa'}$ efficiently. We can rewrite $\mathcal
N_{\kappa\kappa'}$ as the following explicit form
\begin{align}\label{Eq.norm}
  \mathcal{N}_{\kappa\kappa'} = \langle\Phi| a_{1} \cdots  a_{n} [\Omega]  a^\dag_{1'} \cdots a^\dag_{n'} |\Phi \rangle,
\end{align}
which is usually evaluated \cite{PSM-review} by means of the
generalized Wick's theorem that decomposes Eq. (\ref{Eq.norm}) into
a combination of three types of contraction, denoted as $A$, $B$,
and $C$, with their matrix expressions \cite{Wick1979NPA}
\begin{align}
  A_{\nu\nu'}(\Omega) \equiv&\ \langle\Phi| [\Omega] a^\dag_{\nu}a^\dag_{\nu'} |\Phi\rangle = \Big(V^\ast(\Omega)U^{-1}(\Omega)\Big)_{\nu\nu'}, \nonumber \\
  B_{\nu\nu'}(\Omega) \equiv&\ \langle\Phi| a_{\nu}a_{\nu'}[\Omega] |\Phi\rangle = \Big(U^{-1}(\Omega)V(\Omega)\Big)_{\nu\nu'}, \\
  C_{\nu\nu'}(\Omega) \equiv&\ \langle\Phi| a_{\nu}[\Omega]a^\dag_{\nu'} |\Phi\rangle = \Big(U^{-1}(\Omega)\Big)_{\nu\nu'}.  \nonumber
\end{align}
where $U$ and $V$ are the corresponding matrices of action of the
rotation operator on quasiparticles \cite{Wick1979NPA}. 
It was pointed out \cite{Mizusaki2013PLB} that in applying the
generalized Wick's theorem, a matrix element of Eq. (\ref{Eq.norm})
involving $n$ and $n'$ qps, respectively in the left- and right-side
of $[\Omega]$, contains $(n+n'-1)!!$ terms. The number of terms
becomes so large that it is practically impossible to write down
expressions explicitly for more than 4-qp states.

By using the Fermion coherent state and Grassmann integral, a
general expression for the matrix elements (\ref{Eq.norm}) in terms
of the Pfaffian can be derived \cite{Mizusaki2013PLB}
\begin{align}\label{Eq.norm-pf}
  \langle\Phi| a_{1} \cdots  a_{n} [\Omega]  a^\dag_{1'} \cdots a^\dag_{n'} |\Phi \rangle =
  Pf (X) = Pf \left( \begin{array}{cc} B & C \\ -C^T & A \end{array} \right) ,
\end{align}
where $X$ is a skew-symmetric matrix with dimension
$(n+n')\times(n+n')$. The indices of rows and columns for $B$ run
from $1$ to $n$ ($1,\ldots,n$) and the ones for $A$ run from $1'$ to
$n'$ ($1',\ldots,n'$). For matrix $C$ in Eq. (\ref{Eq.norm-pf}), the
indices of rows run from $1$ to $n$ and those of columns run from
$1'$ to $n'$. The Pfaffian is defined as
\begin{align}\label{Eq.pf}
  Pf({\cal A}) \equiv \frac{1}{2^n n!} \sum_{\sigma \in S_{2n}} \text{sgn}(\sigma) \prod^{n}_{i=1} a_{\sigma(2i-1)\sigma(2i)}.
\end{align}
for a skew-symmetric matrix ${\cal A}$ with dimension $2n\times2n$,
of which matrix elements are $a_{ij}$. The symbol $\sigma$ is a
permutation of $\{1,2,3,\ldots,2n\}$, $\text{sgn} (\sigma)$ is its
sign, and $S_{2n}$ represents a symmetry group. 
This makes it possible and efficient to work with the expanded PSM
configuration in (\ref{Eq.config}), since calculations of the
corresponding Pfaffian are not time-consuming \cite{Pf-code-2011}.
Consequently, the new code still runs on a single core of PC
although the configuration space in (\ref{Eq.config}) is now much
extended as compared to that of the original PSM code
\cite{PSM-code}.

The PSM employs the Hamiltonian with separable forces:
\begin{align}\label{Eq.hamil}
\hat H = \hat H_0 -\frac{1}{2}\chi_{QQ} \sum_\mu \hat
Q^\dag_{2\mu}\hat Q_{2\mu} - G_M \hat P^\dag \hat P - G_Q \sum_\mu
\hat P^\dag_{2\mu} \hat P_{2\mu},
\end{align}
where $\hat H_0$ is the spherical single-particle term including the
spin-orbit force \cite{Nil-1985}, and the rest is the
quadrupole+pairing type of separable interactions, which contains
three parts. The strength of the quadrupole-quadrupole term
$\chi_{QQ}$ is determined in a self-consistent manner so that it is
related to deformation of the basis \cite{PSM-review}. The
monopole-pairing strength is taken to be the form $G_M=[G_1 \mp
G_2(N-Z)/A]/A$, where ``$+$" (``$-$") is for protons (neutrons),
with $G_1=20.12$ and $G_2=13.13$ being the coupling constants
\cite{Sun1996PReport}. The quadrupole-pairing strength $G_Q$ is
taken, as usual, to be 16\% of $G_M$ for all the nuclei considered
in this study.

If one keeps axial symmetry in the deformed basis, as it is the case
of the present work, $D^I_{MK}$ in Eq. (\ref{Eq.projec-op}) reduces
to the small $d$-function and the three dimensions in $\Omega$
reduce to one. The following two examples represent first
applications of the extension of the PSM configuration space with
6-qp's by using the Pfaffian algorithm.

\begin{figure}[htbp]
\includegraphics[width=0.42\textwidth]{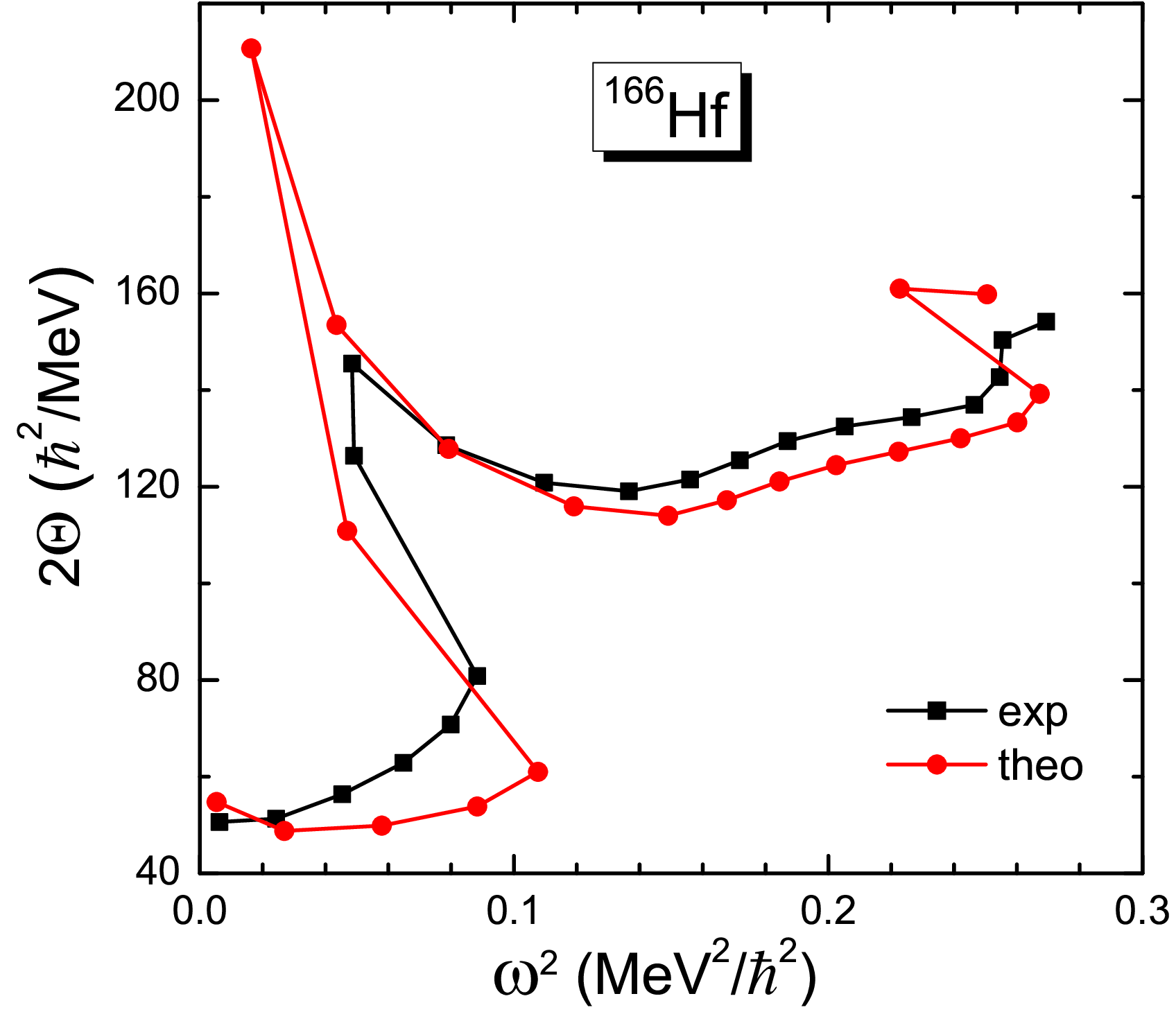}
\caption{\label{fig:hf166a} (Color online) Back-bending plot for
$^{166}$Hf. The calculated results are compared with the
experimental data taken from \cite{Hf166-EPJA2000}.}
\end{figure}

In rotating nuclei, the Coriolis force tends to decouple the
pairing. Nucleons in the highest-$j$ orbital breaks first, leading
to the first anomaly in observed moment of inertia at spin $I
\approx 14$ in rare earth nuclei \cite{First-Backbending}. The
phenomenon is usually displayed in an exaggerated manner with a
back-bending plot, in which twice the moment of inertia $2\Theta$ is
plotted as a function of square of rotational frequency $\omega^2$.
Figure \ref{fig:hf166a} shows the back-bending plot for $^{166}$Hf,
where the theoretical results are compared with the experimental
data. In the calculation, the deformation parameters
$\varepsilon_2=0.208$ and $\varepsilon_4=0.013$ are taken from Ref.
\cite{Moller1995ADNDT}. Anomalies in moment of inertia can be
clearly seen as rotational frequency increases, roughly at $\omega^2
\approx 0.10$, 0.15 and 0.25, corresponding to spin $I \approx 12$,
24 and 34, respectively. The first anomaly exhibits the largest
effect, causing a sharp increase in $2\Theta$ within a small
interval of $\omega^2$. This is known as the first back-bending,
corresponding to breaking and alignment of a neutron $i_{13/2}$
pair. The experimental feature is described by the PSM calculation
although quantitatively the theoretical results show deviations at
low spins and exaggerate the back-bending. The discrepancy that the
calculation shows a more flat curve at low spins could be attributed
to the fact that the present calculation assumes an axially
symmetric potential for deformed single-particle states, while
triaxiality may have an effect on the low-lying states of this mass
region \cite{Chen2013JPG}. The second anomaly in Fig.
\ref{fig:hf166a} corresponds to the small increase in $2\Theta$ at
$\omega^2 \approx 0.15$, which is nicely reproduced by the
calculation. At this rotational frequency, an additional $h_{11/2}$
proton pair is broken and their spins are aligned along the axis of
rotation. The third anomaly belongs to the few known cases that have
ever been observed: $2\Theta$ jumps suddenly again at $\omega^2
\approx 0.25$. The observation is correctly described by the present
calculation, and is understood as a simultaneous breaking of two
neutron $i_{13/2}$ pairs and one $h_{11/2}$ proton pair.

\begin{figure}[htbp]
\includegraphics[width=0.42\textwidth]{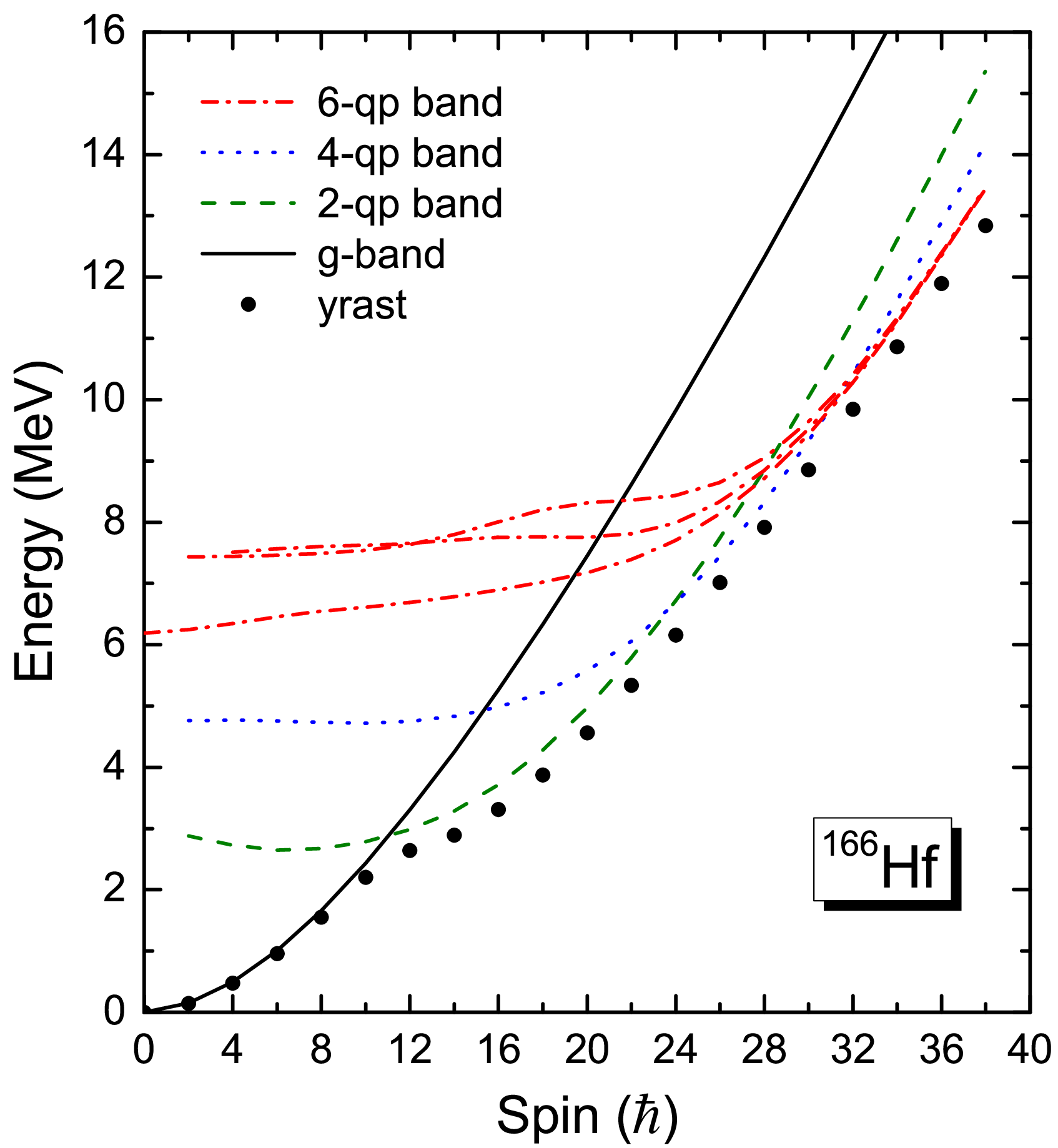}
\caption{\label{fig:hf166b} (Color online) Band diagram for
$^{166}$Hf. Note that only even-spin states are plotted in order to
avoid zigzag in these curves.}
\end{figure}

The calculated results can be analyzed by the so--called band
diagram \cite{PSM-review}, where energies of theoretical bands are
plotted as functions of spin. The energy of a theoretical band
$\kappa$ is defined as
\begin{align}
  E_\kappa (I) = \frac{\langle \Phi_\kappa| \hat H\hat P^I_{KK} |\Phi_\kappa\rangle}{\langle \Phi_\kappa|\hat P^I_{KK} |\Phi_\kappa\rangle}
  ,
\end{align}
which is the projected energy of a multi-qp configuration in
(\ref{Eq.config}). Figure \ref{fig:hf166b} displays the band diagram
for $^{166}$Hf, where the 0-qp ground (g-) band, one 2-qp band, one
4-qp band, and three 6-qp bands are selected from about 200
projected configurations in the calculation because of their
important roles played in the yrast band (marked by dots). It is
seen that the first back-bending at $I \approx 12$ in Fig.
\ref{fig:hf166a} corresponds to the crossing between g-band and the
2-qp (s-) band. The configuration of the s-band is found to be the
neutron 2-qp state $\nu 3/2^+[651] \otimes \nu 5/2^+[642]$ with
$K=1$. The s-band remains to be the yrast band until it is crossed
by a 4-qp band at $I \approx 24 $. This 4-qp band is based on an
addition of an $h_{11/2}$ proton pair, corresponding to the
configuration $\nu 3/2^+[651] \otimes \nu 5/2^+[642] \otimes \pi
7/2^-[523] \otimes \pi 9/2^-[514]$ with $K=2$. Note that due to the
small crossing angle (the ratio between the slopes of two crossing
bands at the crossing point, see Ref. \cite{Hara1991NPA} for
discussion), there is only a slight up-bending in $2\Theta$ at $I
\approx 24$, as seen in Fig. \ref{fig:hf166a}.

The contribution of the new 6-qp configurations in the basis enables
us to understand the third anomaly in moment of inertia at $\omega^2
\approx 0.25$ in Fig. \ref{fig:hf166a}. It is clear that this
anomaly corresponds to the crossing of the 4-qp band with three 6-qp
bands at $I \approx 34$. Two of the 6-qp bands whose energies are
almost the same at low spins have the same configuration $\nu
1/2^+[660] \otimes \nu 3/2^+[651] \otimes \nu 5/2^+[642] \otimes \nu
7/2^+[633] \otimes \pi 7/2^-[523] \otimes \pi 9/2^-[514]$ but with
different $K$ values $K=-1$ and $-3$. The configuration of the third
6-qp band is $\nu 1/2^+[660] \otimes \nu 3/2^+[651] \otimes \nu
5/2^+[642] \otimes \nu 5/2^+[642] \otimes \pi 7/2^-[523] \otimes \pi
9/2^-[514]$ with $K=0$. As the level density increases with spin,
the wave function of the yrast band beyond $I \approx 34$ is found
to have a large admixture of these 6-qp states. On the other hand,
the 6-qp configuration consisting of $\nu (i_{13/2})^2 \nu
(h_{9/2})^2 \pi (h_{11/2})^2$ lies higher in energy, and therefore,
does not cross with the 4-qp band.

\begin{figure*}[htbp]
\includegraphics[width=0.8\textwidth]{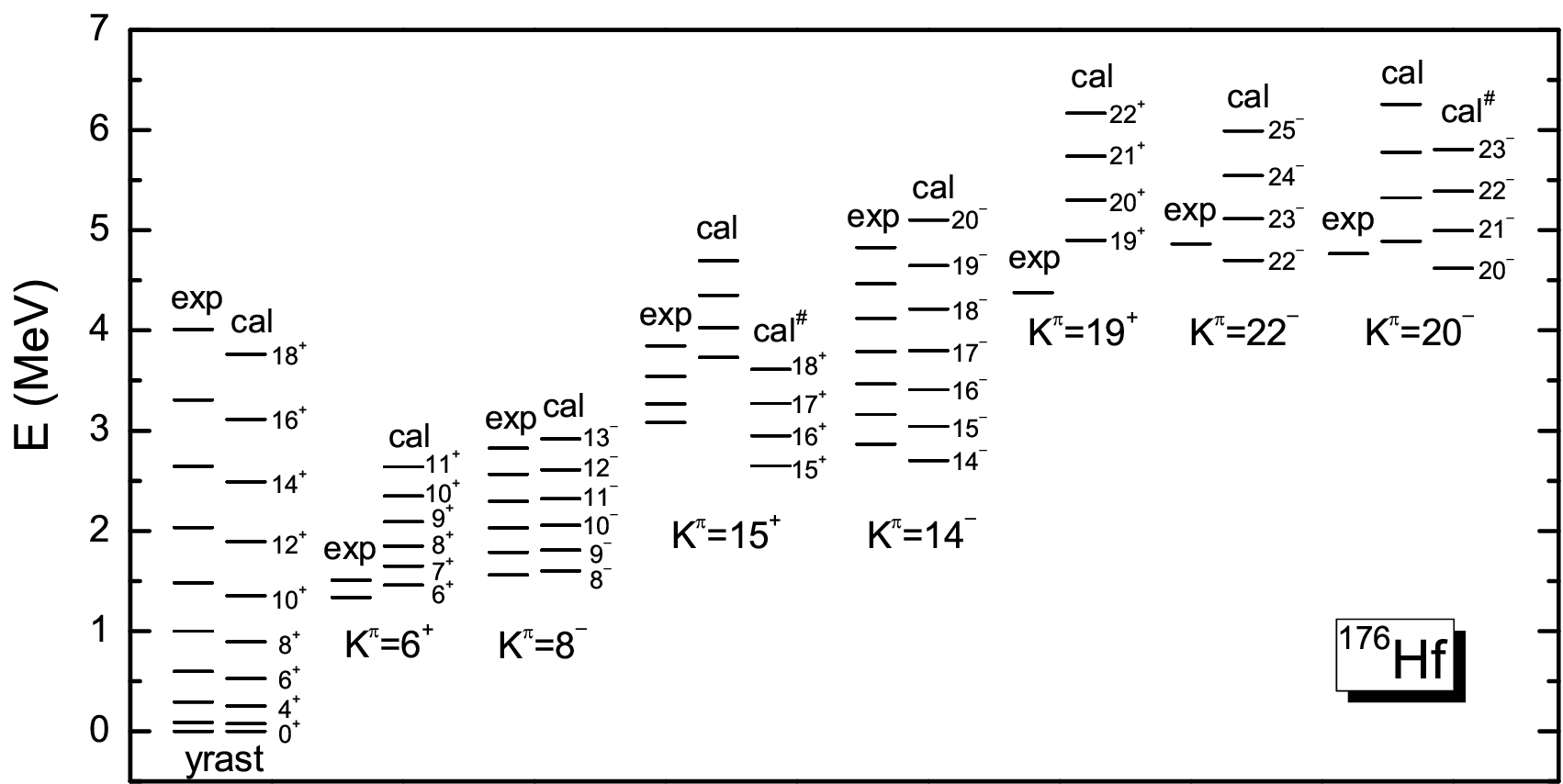}
\caption{\label{fig:hf176} Comparison of the calculated yrast,
2-qp, 4-qp, and 6-qp bands for $^{176}$Hf with available
experimental data taken from \cite{Hf176-2010PRC}.}
\end{figure*}

Our next example to demonstrate the role of 6-qp configurations is
shown in Fig. \ref{fig:hf176} with the calculated multi-qp high-$K$
bands in $^{176}$Hf, compared with the experimental data taken from
Ref. \cite{Hf176-2010PRC}. Several high-$K$ 4-qp and 6-qp isomers in
$^{176}$Hf with higher excitation energies were experimentally known
\cite{Hf176-PRL}. The quadrupole and hexadecapole deformation
parameters for the $^{176}$Hf calculation are adopted as
$\varepsilon_2=0.245$ and $\varepsilon_4=0.024$, which are close to
but slightly different from the deformation parameters in Ref.
\cite{Moller1995ADNDT}. It is seen from Fig. \ref{fig:hf176} that
the yrast band is well reproduced by the PSM calculation. In
consistence with Ref. \cite{Hf176-2010PRC}, our calculation suggests
that the 2-qp $K^\pi=6^+$ and $8^-$ states have $\pi 7/2 [404]
\otimes \pi 5/2[402]$ and $\pi 7/2 [404] \otimes \pi 9/2[514]$,
respectively, as the main configurations for the observed bands.

For the $K^\pi = 15^+$ band, the proposed configuration in Ref.
\cite{Hf176-2010PRC} was a 4-qp $\nu 5/2^-[512] \otimes \nu
9/2^+[624] \otimes \pi 7/2^+[404] \otimes \pi 9/2^-[514]$. Our
calculated band-head energy for this configuration (indicated by
cal) is about 700 keV higher than the corresponding data. The
calculation shows that there is another 4-qp configuration
(indicated by cal$^\#$) with $\nu 7/2^-[514] \otimes \nu 7/2^+[633]
\otimes \pi 7/2^+[404] \otimes \pi 9/2^-[514]$, which is the lowest
one among the same type of configuration (i.e. of the four
constituent quasi-particles originating from the orbitals belonging
to four different major shells). As the band associated with this
configuration lies lower in energy, it should have a larger chance
to be observed in experiment.

The rotational band built on the $K^\pi = 14^-$, 2866-keV isomer was
observed up to $I=20$ \cite{Hf176-PRL}. One can see from Fig.
\ref{fig:hf176} that both the excitation energy of this isomer and
the corresponding rotational band are well reproduced by the
calculation. The configuration is found to be the 4-qp $\nu
5/2^-[512] \otimes \nu 7/2^-[514] \otimes \pi 7/2^+[404] \otimes \pi
9/2^-[514]$, consistent with the one suggested in Ref.
\cite{Hf176-2010PRC}. Nevertheless, the calculated moment of inertia
is smaller than the corresponding experimental value, i.e., the
calculated level spacings between adjacent levels of the band is
larger than the experimental ones. The same problem seems to exist
in the above-discussed $K^\pi = 15^+$ 4-qp band. Such a discrepancy
could be due to the fact that pairing correlations are treated by
the BCS method in the present model, which does not account for the
blocking effects properly.

At least three 6-qp structures have been known in $^{176}$Hf
\cite{Hf176-2010PRC}: the 4377-keV $K^\pi = 19^+$ and 4864-keV
$K^\pi = 22^-$ isomers, and the 4767-keV $K^\pi = 20^-$ state. In
Fig. \ref{fig:hf176}, we show the predicted rotational bands based
on these configurations. The calculation suggests that the band
built on the $K^\pi = 19^+$ isomer has the main configuration $\nu
1/2 [521] \otimes \nu 5/2 [512] \otimes \nu 7/2 [514] \otimes \nu
9/2 [624] \otimes \pi 7/2 [404] \otimes \pi 9/2[514]$, consistent
with the assignment given in Ref. \cite{Hf176-2010PRC}. The 6-qp
isomer with $K^\pi = 22^-$ was assigned to be of the configuration
$\nu 5/2^-[512] \otimes \nu 7/2^-[514] \otimes \nu 7/2^+[633]
\otimes \nu 9/2^+[624] \otimes \pi 7/2^+[404] \otimes \pi
9/2^-[514]$. The experimental energy of this isomer is well
reproduced by the PSM. This isomer is found to decay to the $K^\pi =
20^-$ state by $E2$ transition \cite{Hf176-2010PRC}. The
configuration of the $K^\pi = 20^-$ state was assigned to be $\nu
1/2^-[521] \otimes \nu 7/2^-[514] \otimes \nu 7/2^+[633] \otimes \nu
9/2^+[624] \otimes \pi 7/2^+[404] \otimes \pi 9/2^-[514]$
\cite{Hf176-2010PRC}. The calculated excitation energy for this
configuration is found to be slightly higher than another 6-qp
configuration $\nu 5/2^-[512] \otimes \nu 7/2^-[514] \otimes \nu
5/2^+[642] \otimes \nu 7/2^+[633] \otimes \pi 7/2^+[404] \otimes \pi
9/2^-[514]$ (indicated by cal$^\#$).

\begin{table}[htbp]
\caption{\label{tab:BE2} Comparison of calculated $B(E2)$ values (in
W.u.) for the isomeric states in $^{176}$Hf with the available data
taken from \cite{Hf176-BE2-data, Hf176-2010PRC}.}
\begin{ruledtabular}
\begin{tabular}{cccccc}
$K^\pi_i$ & $I_i$ & $K^\pi_f$ & $I_f$ & \multicolumn{2}{c}{$B(E2;
I_i \rightarrow I_f)$}  \\ \cline{5-6}
          &       &           &       &         exp.         &         cal.          \\  \hline
$6^+$     &   6   &  $0^+$    &  6    & $2.82\times 10^{-6}$ & $1.93\times 10^{-9}$  \\
$6^+$     &   6   &  $0^+$    &  4    & $3.16\times 10^{-7}$ & $2.21\times 10^{-9}$  \\
$14^-$    &   14  &  $8^-$    &  12   & $5.33\times 10^{-7}$ & $5.55\times 10^{-8}$  \\
$22^-$    &   22  &  $20^-$   &  20   & $4.8 \times 10^{-3}$ & $3.7 \times 10^{-2}$  \\
\end{tabular}
\end{ruledtabular}
\end{table}

Calculated $E2$ transition probabilities associated with isomer
decay are compared with available experimental data in Table I. The
first three transitions in Table I are $K$-forbidden ones with
$\Delta K=6$. Our calculated results are 1-3 orders of magnitude
smaller than the data. Again, we attribute such a discrepancy to the
absence of triaxial degree of freedom in the deformed basis. In Ref.
\cite{Chen2013JPG}, it was demonstrated that the $K$-forbidden
transition from the $6^+$ 2-qp isomer to the ground-state band is
sensitive to mixing with the $6^+$ state of the $\gamma$-vibrational
band, which could be accounted for when a triaxial deformed
single-particle basis with three-dimensional angular momentum
projection is employed in the model. On the other hand, the
calculated $B(E2)$ for the allowed transition from the $22^-$ isomer
to the $20^-$ state is $3.7\times 10^{-2}$ W.u., in a reasonable
agreement with the experimental value $4.8\times 10^{-3}$ W.u.
obtained in \cite{Hf176-2010PRC}.

In summary, the Pfaffian algorithm has been recently introduced to
facilitate computer codings in realistic shell-model calculations.
By using the Pfaffian algorithm for computing overlap matrix
elements, the configuration space of the PSM has been expanded, for
the first time, to include all kinds of 4-qp and some 6-qp states
for both positive and negative parities. As an initial application,
contributions of the 4-qp and 6-qp states in the yrast band at high
spins have been analyzed. It is found that the third anomaly in
moment of inertia at spin $I \approx 34$ in $^{166}$Hf could be
explained as the band crossing with the 6-qp bands. Multi-qp
high-$K$ isomers in $^{176}$Hf have also been investigated, where
the experimentally observed high-$K$ isomers at high excitation
energies have been described as various 2-, 4-, and 6-qp
configurations.

The present work has been restricted in the axial symmetric case in
the deformed basis. The current studies of the Triaxial PSM can only
afford to use a very small multi-qp configuration space
\cite{Chen2013JPG, Sheikh-TPSM}. It remains to be seen how the
Pfaffian algorithm can help to simplify the calculation with angular
momentum projection in a three-dimensional space.

Valuable discussions with Z.-C. Gao and Q.-L. Hu are acknowledged.
Research at SJTU was supported by the National Natural Science
Foundation of China (No. 11135005) and by the 973 Program of China
(No. 2013CB834401).



\begin{thebibliography}{99}
%
\bibitem{Mottelson-Valatin}
B. R. Mottelson and J. G. Valatin, Phys. Rev. Lett. {\bf 5}, 511
(1960).

\bibitem{Ring}
U. Mutz and P. Ring, J. Phys. G: Nucl. Phys. {\bf 10}, L39 (1984).

\bibitem{Stephens-Simon}
F. S. Stephens and R. S. Simon, Nucl. Phys. A {\bf 183}, 257 (1972).

\bibitem{First-Backbending}
A. Johnson, H. Ryde, and J. Sztarkier, Phys. Lett. {\bf 34B}, 605
(1971); A. Johnson, H. Ryde, and S. A. Hjorth, Nucl. Phys. A {\bf
179}, 753 (1972).

\bibitem{Second-Backbending}
I. Y. Lee {\it et al.}, Phys. Rev. Lett. {\bf 38}, 1454 (1977).

\bibitem{Third-Backbending}
J. Burde {\it et al.}, Phys. Rev. Lett. {\bf 48}, 530 (1982).

\bibitem{Hf166-EPJA2000}
D. R. Jensen \emph{et al}., Eur. Phys. J. A \textbf{8}, 165 (2000).

\bibitem{Hf168PRC2009}
R. B. Yadav \emph{et al}., Phys. Rev. C \textbf{80}, 064306 (2009).

\bibitem{Walker1999Nature}
P. M. Walker and G. D. Dracoulis, Nature \textbf{399}, 35 (1999).

\bibitem{Sun2005Nat-Phys}
A. Aprahamian and Y. Sun, Nat. Phys. \textbf{1}, 81 (2005).

\bibitem{Walker2005PhysTod}
P. M. Walker and J. J. Carroll, Phys. Today \textbf{58} (2005).

\bibitem{Xu2004PRL}
F. R. Xu, E. G. Zhao, R. Wyss, and P. M. Walker, Phys. Rev. Lett.
\textbf{92}, 252501 (2004).

\bibitem{Herzberg2006Nature}
R. D. Herzberg \emph{et al.}, Nature \textbf{442}, 896 (2006).

\bibitem{Purry95} C. S. Purry {\it et al.}, Phys. Rev. Lett. {\bf
   75}, 406 (1995).
\bibitem{Dra05} G. D. Dracoulis {\it et al.}, Phys. Rev. C {\bf
   71}, 044326 (2005).
\bibitem{Reed10} M. W. Reed {\it et al.}, Phys. Rev. Lett. {\bf
   105}, 172501 (2010).

\bibitem{Ring2004many-body}
P. Ring and P. Schuck, \emph{The nuclear many-body problem}
(Springer Verlag, 2004).

\bibitem{PSM-review}
K. Hara and Y. Sun, Int. J. Mod. Phys. E \textbf{4}, 637 (1995).

\bibitem{Chen2012PRC} F.-Q. Chen, Y.-X. Liu, Y. Sun, P. M. Walker, and G. D.
   Dracoulis, Phys. Rev. C {\bf 85}, 024324 (2012).

\bibitem{Liu2011NPA} Y.-X. Liu, Y. Sun, X.-H. Zhou, Y.-H. Zhang, S.-Y. Yu,
Y.-C. Yang, H. Jin, Nucl. Phys. A {\bf 858}, 11 (2011).

\bibitem{Robledo2009PRC} L. M. Robledo, Phys. Rev. C {\bf 79}, 021302(R) (2009).

\bibitem{Onishi1996NP}
N. Onishi and S. Yoshida, Nucl. Phys. {\bf 80}, 367 (1966).

\bibitem{Robledo2011PRC}
L. M. Robledo, Phys. Rev. C {\bf 84}, 014307 (2011).

\bibitem{Robledo2012PRL} G. F. Bertsch and L. M. Robledo,
Phys. Rev. Lett. {\bf 108}, 042505 (2012).

\bibitem{Bender2012PRC}
B. Avez and M. Bender, Phys. Rev. C {\bf 85}, 034325 (2012).

\bibitem{Oi-Misuzaki2012PLB}
M. Oi and T. Mizusaki, Phys. Lett. B {\bf 707}, 305 (2012).

\bibitem{Misuzaki-Oi2012PLB}
T. Mizusaki and M. Oi, Phys. Lett. B {\bf 715}, 219 (2012).

\bibitem{Hu2013}
Q.-L. Hu, Z.-C. Gao, and Y. S. Chen, Phys. Lett. B {\bf 734}, 162
(2014).

\bibitem{Mizusaki2013PLB}
T. Mizusaki, M. Oi, F. Q. Chen, and Y. Sun, Phys. Lett. B
\textbf{725}, 175 (2013).

\bibitem{Nilsson1969}
S. G. Nilsson \emph{et al}., Nucl. Phys. A \textbf{131}, 1 (1969).

\bibitem{Angular-book}
D. A. Varshalovich, A. N. Moskalev, and V. K. Khersonskii,
\emph{Quantum theory of angular momentum} (World Scientific, 1988).

\bibitem{Wick1979NPA}
K. Hara and S. Iwasaki, Nucl. Phys. A \textbf{332}, 61 (1979).

\bibitem{Pf-code-2011}
C. Gonz\'alez-Ballestero, L. M. Robledo, and G. F. Bertsch, Comput.
Phys. Commun. \textbf{182}, 2213 (2011).

\bibitem{PSM-code}
Y. Sun and K. Hara, Comput. Phys. Commun. \textbf{104}, 245 (1997).

\bibitem{Nil-1985}
T. Bengtsson and I. Ragnarsson, Nucl. Phys. A \textbf{436}, 14
(1985).

\bibitem{Sun1996PReport}
Y. Sun and D. H. Feng, Phys. Rep. \textbf{264}, 375 (1996).

\bibitem{Moller1995ADNDT}
P. Moller, J. R. Nix, W. D. Myers, and W. J. Swiatecki, At. Data
Nucl. Data Tables \textbf{59}, 185 (1995).

\bibitem{Chen2013JPG}
F.-Q. Chen, Y. Sun, P. M. Walker, G. D. Dracoulis, Y. R. Shimizu,
and J. A. Sheikh, J. Phys. G: Nucl. Part. Phys. {\bf }40, 015101
(2013).

\bibitem{Hara1991NPA}
K. Hara and Y. Sun, Nucl. Phys. A {\bf 529}, 445 (1991).

\bibitem{Hf176-2010PRC}
G. Mukherjee \emph{et al}., Phys. Rev. C \textbf{82}, 054316 (2010).

\bibitem{Hf176-PRL}
T. L. Khoo, F. M. Bernthal, R. G. H. Robertson, and R. A. Warner,
Phys. Rev. Lett. \textbf{37}, 823 (1976).

\bibitem{Hf176-BE2-data}
M. S. Basunia, Nucl. Data Sheets {\bf 107}, 791 (2006).

\bibitem{Sheikh-TPSM}
J. A. Sheikh, G. H. Bhat, Y.-X. Liu, F.-Q. Chen, Y. Sun, Phys. Rev.
C {\bf 84}, 054314 (2011); J. A. Sheikh, G. H. Bhat, Y. Sun, R.
Palit, Phys. Lett. B {\bf 688}, 305 (2010).

%
\end{thebibliography}
\end{document}